\documentclass[amsmath, amsfonts, showpacs, superscriptaddress, twocolumn, prb]{revtex4}
\usepackage{graphicx}
\usepackage{epsfig}
\usepackage{bm}
\usepackage{dcolumn}
\usepackage{amsmath}
\usepackage{amssymb}
\usepackage{color}
\usepackage{stackrel}
\usepackage{accents}
\usepackage{latexsym}
\usepackage{verbatim}
\usepackage{hyperref}
\usepackage{bbm}

\newcommand{\beg}{\begin{equation}}
\newcommand{\en}{\end{equation}}

\newcommand{\bp}{\mathbf p}

\newcommand{\bk}{\mathbf k}
\newcommand{\br}{\mathbf r}

\newcommand{\bR}{\mathbf R}

\newcommand{\veps}{\varepsilon}

\newcommand{\up}{\uparrow}
\newcommand{\dn}{\downarrow}
\newcommand{\dg}{^\dagger}

\newcommand{\av}[1]{\langle#1\rangle}

\begin{document}

\title{Quasiclassical theory of $C_4$-symmetric magnetic order in disordered multiband metals}

\author{Maxim Dzero}
\affiliation{Department of Physics, Kent State University, Kent, Ohio 44242, USA}

\author{Maxim Khodas}
\affiliation{The Racah Institute of Physics, The Hebrew University of Jerusalem, Jerusalem 9190401, Israel}

\vspace{10pt}

\begin{abstract} 
Recent experimental studies performed in the normal state of iron-based superconductors have discovered the existence of the $C_4$-symmetric (tetragonal) itinerant magnetic state. This state can be described as a spin density wave with two distinct magnetic vectors ${\vec Q}_1$ and ${\vec Q}_2$. Given an itinerant nature
of magnetism in iron-pnictides, we develop a quasiclassical theory of tetragonal magnetic order in disordered three-band metal with
anisotropic band structure. Within our model we find that the $C_4$-symmetric magnetism competes with the $C_2$-symmetric state with a single ${\vec Q}$ magnetic structure vector. Our main results is that disorder promotes tetragonal magnetic state which is in agreement with earlier theoretical studies. 
\end{abstract}


\maketitle

\section{Introduction}
Quasiclassical approach to interacting many-body systems has proved to be a powerful tool in describing their transport and thermodynamic properties. Within this method, the quantum mechanical averages of an operator corresponding to a physical quantity are replaced with the averages of its classical counterpart over all classical trajectories. Alternatively, one can formulate the quasiclassical theory by using the quasiclassical functions which are obtained from  the quantum mechanical single-particle propagators by integrating them over all single particle energies. Qualitatively, for a superconductor with pairing gap $\Delta$ and quasiparticles with Fermi momentum $p_F$ and Fermi velocity $v_F$, this procedure corresponds to averaging over the short length scales of the problem 
$\sim p_F^{-1}$ and retaining the physics at long scales $\sim v_F/\Delta$. Quasiclassical theory was particularly useful in the comparatively recent analysis of the problem of far-from-equilibrium order parameter dynamics in charge-neutral superfluids. \cite{Barankov2004,Emil2005a,Emil2005b,Emil2006,Big-Quench-Review2015}

Most recently, several nontrivial phenomena have been observed in a family of iron-based superconductors and their alloys.\cite{Matsuda_review2014} One example of such phenomena is an observation of the peak in the penetration depth in $\mathrm{BaF}{\mathrm{e}}_{2}{(\mathrm{A}{\mathrm{s}}_{1\ensuremath{-}x}{\mathrm{P}}_{x})}_{2}$ as a function of phosphorus concentration \cite{Carrington_BaFeAs2010,CoExistExp2,Matsuda_Science2012,Auslaender2015}, in ${\mathrm{Ba}}_{1\ensuremath{-}x}$${\mathrm{K}}_{x}$${\mathrm{Fe}}_{2}$${\mathrm{As}}_{2}$ as a function of potassium concentration \cite{CoExistExp3} and, most recently in $\textrm{Ba}(\textrm{Fe}_{1-x}\textrm{Co}_x)_2\textrm{As}_2$ as a function of cobalt concentration.\cite{LondonCobalt} Another example is the experimental observation of the spin-density-wave order which is characterized by two magnetic ordering vectors, ${\vec Q}_1$ and ${\vec Q}_2$, in various alloys iron-based superconducting alloys.\cite{DoubleQExp1,DoubleQExp2,DoubleQExp3,DoubleQExp4,DoubleQExp5,DoubleQExp6,DoubleQExp7,DoubleQExp8}

Due to the fact that in iron-based superconductors the superconductivity is often observed near magnetic instability, quasiclassical approaches initially developed for the purely superconducting states have been re-formulated to specifically include the effects of competition between superconducting and magnetic phases as well as the effects of disorder.\cite{Fernandes-Tc,Vavilov2011,Dzero2015,SNS2019} The experimental observations of the peak in the London penetration depth remains only partially understood \cite{Sasha,Sachdev} which provides an additional motivation to look for possible explanations of this effect.

In turn, the experimental discovery of the double-${\vec Q}$ magnetic state in iron-based superconductors has lead to an appearance of many theoretical works discussing the emergence of this state and its various properties as well as its relation with other magnetic states.\cite{DQTheory1,DQTheory2,DQTheory3,DQTheory4,DQTheory5,DQTheory6,DQTheory7,DQTheory8,DQTheory9}
Most recently, the effects of disorder on the stability of the single- and double-${\vec Q}$ states have been discussed.\cite{AlexC4} In particular, it was found that disorder leads to suppression of the single-{$\vec Q$} state in favor the the double-${\vec Q}$ one. 

Inspired by the earlier work on this problem, 
in this paper we use a slightly simplified version of the model introduced in Ref. [\onlinecite{AlexC4}] to formulate a quasiclassical theory of the double-${\vec Q}$ state in iron-based superconductors. Specifically, we consider the disordered model which incorporates both interband and intraband disorder. In agreement with the earlier results \cite{AlexC4}, we find that when the interband disorder can be ignored, the intraband disorder promotes the emergence of the double-${\vec Q}$ state. 

This paper is organized as follows. In the next Section II introduce the model Hamiltonian. Section III is devoted to the formulation of the quasiclassical approach with the derivation of the quasiclassical equations. In Section IV contains the results of the Landau expansion for the free energy using the quasiclassical equations. Section V contains the discussion of the results and comments related to the further development of the presented formalism in the context of the physics of iron-based superconductors. Sections with acknowledgements and Appendix with some technical details conclude the paper.

\section{Model}
In what follows we first introduce the model Hamiltonian, which consists of three terms:
\beg\label{Eq1}
\hat{H}=\hat{H}_0+\hat{H}_{\textrm{sdw}}+\hat{H}_{\textrm{dis}}.
\en
The first term on the right hand side of this expression is a single-particle Hamiltonian which describes the band-structure consisting of three bands: 
hole-like band at the $\Gamma$ point and two electron-like bands 
centered at ${\vec Q}_X=(\pi,0)$, ${\vec Q}_Y=(0,\pi)$ of the two-dimensional Brillouin zone. We use the compact notations to write down $\hat{H}_0$ using the six-component spinor 
$\hat{\Psi}_\bk\dg=\left(\hat{c}_{\bk\up}\dg, ~\hat{c}_{\bk\dn}\dg, ~\hat{d}_{\bk\up}\dg, ~\hat{d}_{\bk\dn}\dg,
~\hat{f}_{\bk\up}\dg, ~\hat{f}_{\bk\dn}\dg\right)$:
\beg\label{H0}
\hat{H}_0=\sum\limits_{\bk}\hat{\Psi}_\bk\dg\left(
\begin{matrix}
\veps_\Gamma(\bk)\hat{\sigma}_0 & 0 & 0 \\
0 & \veps_X(\bk)\hat{\sigma}_0 & 0 \\
0 & 0 & \veps_Y(\bk)\hat{\sigma}_0
\end{matrix}
\right)\hat{\Psi}_\bk,
\en
where $\hat{\sigma}_0$ is a unit $2\times 2$ matrix and single particle energy spectra are given by 
$\veps_\Gamma(\bk)=-\xi_\bk$, $\xi_\bk=\epsilon_0-{k^2}/{2}$, $\veps_X(\bk)=\xi_\bk+\delta_0+\delta_2\cos2\phi$, 
$\veps_Y(\bk)=\xi_\bk+\delta_0-\delta_2\cos2\phi$, $\epsilon_0$ is the energy which amounts to the off-set between the bands and ${\mathbf k}=(k\cos\phi,k\sin\phi)$. Here $\delta_0$ is an anisotropy parameter which is defined relative to the chemical potential $\mu$, so that the bands are perfectly nested when $\delta_0=0$. Lastly, $\delta_2$ is an anisotropy parameter which accounts for the ellipticity of the corresponding Fermi pockets. \cite{AlexC4}

The second term, $\hat{H}_{\textrm{sdw}}$, appearing in (\ref{Eq1}) accounts for the spin-density-wave order within the mean-field approximation:
\beg\label{Hsdw}
\hat{H}_{\textrm{sdw}}=-\sum\limits_{\bk}\hat{\Psi}_\bk\dg\left(
\begin{matrix}
0 & {\vec m}_X\cdot\vec{\sigma} & {\vec m}_Y\cdot\vec{\sigma} \\
{\vec m}_X\cdot\vec{\sigma} & 0 & 0 \\
{\vec m}_Y\cdot\vec{\sigma} & 0 & 0
\end{matrix}
\right)\hat{\Psi}_\bk.
\en
Here ${\vec m}_X$, ${\vec m}_Y$ are the magnetizations corresponding to two structure vectors ${\vec Q}_X$ and ${\vec Q}_Y$. In what follows, we will assume that magnetic state has Ising-like anisotropy, so we replace 
${\vec m}_{X,Y}\cdot\vec{\sigma}\to m_{X,Y}\hat{\sigma}_3$. Within the mean-field approach we have adopted here, the order parameters $m_{X,Y}$ must be computed self-consistently.

Finally, the last term on the r.h.s. side of Eq. (\ref{Eq1}) introduces the disorder potential in a system. In principle, the disorder should scatter quasiparticles within each band (intraband scattering) as well as between the bands (interband scattering). The disorder unavoidably leads to the suppression of itinerant magnetism. In this paper we will limit ourselves to the case of an intraband disorder only, for an interband disorder scattering only plays a crucial role in the problem of co-existence of magnetism and superconductivity,\cite{Vavilov2011,Fernandes-Tc,Dzero2015} while for the problem at hand it will only lead the faster suppression of the magnetic order. Thus, we write for the last term in (\ref{Eq1})
\beg\label{Hdis}
\hat{H}_{\textrm{dis}}=u\int d^2\br{\Psi}\dg(\br){\Psi}(\br)\sum\limits_i\delta(\br-\bR_i)
\en
and the summation is performed over the impurity sites. 

\section{Quasiclassical equations}
In order to formulate the quasiclassical theory, we first introduce a single-particle correlation function 
\beg\label{Gtwo}
G_{\alpha\beta}(x,x')=
-\left\langle\hat{T}_\tau\left(\hat{\Psi}_\alpha(x)\hat{\Psi}_\beta\dg(x')\right)\right\rangle_{\textrm{g.s.}}
\en
in the Matsubara representation, $\hat{\Psi}_\alpha(x)=\hat{\Psi}_\alpha(\br,\tau)$, and the averaging is performed over the ground state of the Hamiltonian, Eq. (\ref{Eq1}). Next step consists in employing the equations of motion for the propagator (\ref{Gtwo}):
\beg\label{Eqs1}
\begin{split}
-\frac{\partial}{\partial \tau}\hat{G}-\hat{H}_\br\hat{G}-\hat{\Sigma}\circ\hat{G}&=\delta(x-x')\mathbbm{1}, \\
\frac{\partial}{\partial \tau'}\hat{G}-\hat{G}\hat{H}_{\br'}-\hat{G}\circ\hat{\Sigma}&=\delta(x-x')\mathbbm{1}.
\end{split}
\en
Here $\hat{H}_\br$ acts on $\br$, the self-energy part $\hat{\Sigma}$ is generated by the disorder potential and its action on the propagator is
\beg\label{circ}
\hat{\Sigma}\circ\hat{G}=\int\limits_0^{1/T}d\tau''\int d^2\br''\Sigma_{\alpha\gamma}(x,x'')G_{\gamma\beta}(x'',x').
\en
The summation over the repeated indices is assumed. Next, we perform the Wigner transformation
\beg\label{Wigner}
\hat{G}(x,x')=\int\frac{d^2\bk}{(2\pi)^2}e^{i\bk\cdot(\br-\br')}\hat{G}\left(\tau-\tau';\frac{\br+\br'}{2},\bk\right).
\en
In the presence of the quenched disorder, propagators will be dependent on $\bR=(\br+\br')/2$. In what follows we assume that the disorder in uncorrelated and will average the propagator over the disorder distribution which corresponds to self-consistent Born approximation. Lastly, we introduce the following matrices:
\beg\label{M111}
\begin{split}
\hat{\cal M}_{1}&=\left(
\begin{matrix}
0 & 0 & 0 \\
0 & \hat{\sigma}_0 & 0  \\
0 & 0 & \hat{\sigma}_0
\end{matrix}
\right), ~\hat{\cal M}_{2}=\left(
\begin{matrix}
0 & 0 & 0 \\
0 & \hat{\sigma}_0 & 0  \\
0 & 0 & -\hat{\sigma}_0
\end{matrix}
\right), \\ 
\hat{\cal M}_{3}&=\left(
\begin{matrix}
\hat{\sigma}_0 & 0 & 0 \\
0 & -\hat{\sigma}_0 & 0  \\
0 & 0 & -\hat{\sigma}_0
\end{matrix}
\right), ~\hat{\cal P}_{X}=\left(
\begin{matrix}
0 & \hat{\sigma}_3 & 0 \\
-\hat{\sigma}_3 & 0 & 0  \\
0 & 0 & 0
\end{matrix}
\right), \\
\hat{\cal P}_{Y}&=\left(
\begin{matrix}
0 & 0 & \hat{\sigma}_3 \\
0 & 0 & 0  \\
 -\hat{\sigma}_3 & 0 & 0
\end{matrix}
\right), ~\hat{\cal Q}_X=\left(
\begin{matrix}
0 & 0 & 0 \\
0 & 0 & \hat{\sigma}_0  \\
0 & \hat{\sigma}_0 & 0
\end{matrix}
\right).
\end{split}
\en
Quasiclassical equations can now be derived after we multiply the first equation (\ref{Eqs1}) from the left and the second equation from the right by $\hat{\cal M}_{3}$. Subtracting the resulting first equation from the second one we find
\beg\label{Quasi1}
\begin{split}
&\left[\omega_n\hat{\cal M}_3,\hat{\cal G}(i\omega_n,\phi_\bp)\right]+i\delta_0\left[\hat{\cal M}_1,\hat{\cal G}(i\omega_n,\phi_\bp)\right]\\&+
i\delta_2\cos(2\phi_\bp)\left[\hat{\cal M}_2,\hat{\cal G}(i\omega_n,\phi_\bp)\right]\\&+i
\left[\left(\hat{H}_\textrm{sdw}+\hat{\Sigma}_{\textrm{dis}}(i\omega_n)\right)\hat{\cal M}_3,\hat{\cal G}(i\omega_n,\phi_\bp)\right]=0, 
\end{split}
\en
where we introduced the quasiclassical function, $[\hat{f},\hat{g}]$ implies the usual commutation relation and
\beg\label{quasi}
\hat{\cal G}(i\omega_n,\phi_\bp)=\frac{i}{\pi}\int {\hat{\cal M}}_3\hat{G}(i\omega_n,\bp)d\xi_\bp.
\en
The self-energy part is determined by the quasiclassical function and disorder scattering rate 
$\Gamma=\pi\nu_F|u|^2$ ($\nu_F$ is the density of states at the Fermi level per valley per spin):
\beg\label{selfe}
\hat{\Sigma}_{\textrm{dis}}(i\omega)=-i{\Gamma}\int\limits_0^{2\pi}\frac{d\phi}{2\pi}\hat{\cal M}_3\hat{\cal G}(i\omega,i\phi_\bp).
\en 
Quasiclassical equation (\ref{Quasi1}) is linear in $\hat{\cal G}$ and therefore is not sufficient to find $\hat{\cal G}$ unambiguously. In order to define the problem completely, one has to complement (\ref{quasi}) with a certain constraint. To derive this constraint, we introduce a new (matrix) function \cite{Aleiner2006} 
\beg\label{B}\nonumber
\hat{\cal B}(\tau,\tau';\phi_\bp)=\int\limits_0^{1/T}\hat{\cal G}(\tau,\tau'';\phi_\bp)\hat{\cal G}(\tau'',\tau';\phi_\bp)d\tau''.
\en
Equation for this matrix function can be easily derived from (\ref{Quasi1}). It then follows that quasiclassical functions must satisfy the following normalization condition:
\beg\label{norm}
\hat{\cal G}^2(i\omega_n,\phi_\bp)=
\mathbbm{1}.
\en
In order to solve the quasiclassical equations (\ref{Quasi1}) self-consistently, we need to specify the matrix structure of the function $\hat{\cal G}$. 
\subsection{Clean system}
We start by setting the disorder scattering rate to zero, $\Gamma=0$, for it would allow us to keep the resulting expressions more compact. Most of the results derived in this Section are easily generalized for the case  when $\Gamma\not=0$ (see below).

In the absence of the magnetic order, the expression for the function $\hat{\cal G}$ follows from (\ref{quasi}) by comparing the solution of the quasiclassical equations with the expression found from the expression for the single-particle propagator, so that a term proportional to $\hat{\cal M}_3$ must appear in the expression for $\hat{\cal G}$. This conjecture also implies that there should also appear two other terms proportional to $\hat{\cal M}_1$ and $\hat{\cal M}_2$ so we write the following ansatz
\beg\label{G0}
\hat{\cal G}_0=\frac{1}{2}(g_1+g_2)\hat{\cal M}_1+\frac{1}{2}(g_2-g_1)\hat{\cal M}_2+g_3\hat{\cal M}_3.
\en
The commutators which include $\hat{H}_{\textrm{sdw}}$ must lead to the appearance of the three more terms in $\hat{\cal G}$: each one of the two of them being proportional to the corresponding magnetizations, while the third one being proportional to the product of $m_X$ and $m_Y$. The calculation yields the following expression
\beg\label{Gclean}
\hat{\cal G}-\hat{\cal G}_0=p_x\hat{\cal P}_X+p_y\hat{\cal P}_Y+q_x\hat{\cal Q}_X.
\en
After plugging this ansatz into the quasiclassical equations and collecting the terms proportional to the 
same matrices (these matrices are different from those introduced above and will not be listed here), we derive the following set of quasiclassical equations:
\beg\label{sol}
\begin{split}
&\left[2i\Omega_n+\delta_2\cos(2\phi)\right]p_x+m_X(g_2-2g_3)=-m_Yq_x, \\
&\left[2i\Omega_n-\delta_2\cos(2\phi)\right]p_y+m_Y(g_1-2g_3)=-m_Xq_x, \\
&2\delta_2\cos(2\phi)q_x=m_Yp_x-m_Xp_y
\end{split}
\en
and $\Omega_n=\omega_n-i\delta_0/2$. Furthermore, given the expression
(\ref{Gclean}) the constraint condition (\ref{norm}) reduces to the set of the following simple relations:
\beg\label{constraint}
\begin{split}
&q_x=-g_1\frac{p_y}{p_x}=-g_2\frac{p_x}{p_y}, ~(g_3-g_1-g_2)^2=1, \\
&p_x^2=g_1(2g_3-g_1-g_2), \quad p_y^2=g_2(2g_3-g_1-g_2).
\end{split}
\en
Note, that by combining the first two relations with the last two ones one also finds $q_x^2=g_1g_2$.
With the help of relations (\ref{constraint}) it is also straightforward to show that the third equation in (\ref{sol}) is redundant, so overall we have got the system of six non-linear equations with six unknowns. These equations must also be supplemented by the self-consistency conditions for the magnetizations, which in terms of the quasiclassical functions have the following form:
\beg\label{selfcons}
m_{X,Y}=-2\pi\nu_Fg_{\textrm{sdw}}T\textrm{Im}\sum\limits_{\omega_n>0}\av{p_{x,y}(i\omega_n,\phi_\bk)},
\en
where $\av{f}$ denotes averaging over $\phi_\bk$ and $g_{\textrm{sdw}}$ is the coupling constant. 

\begin{figure}[h]
\includegraphics[width=8cm]{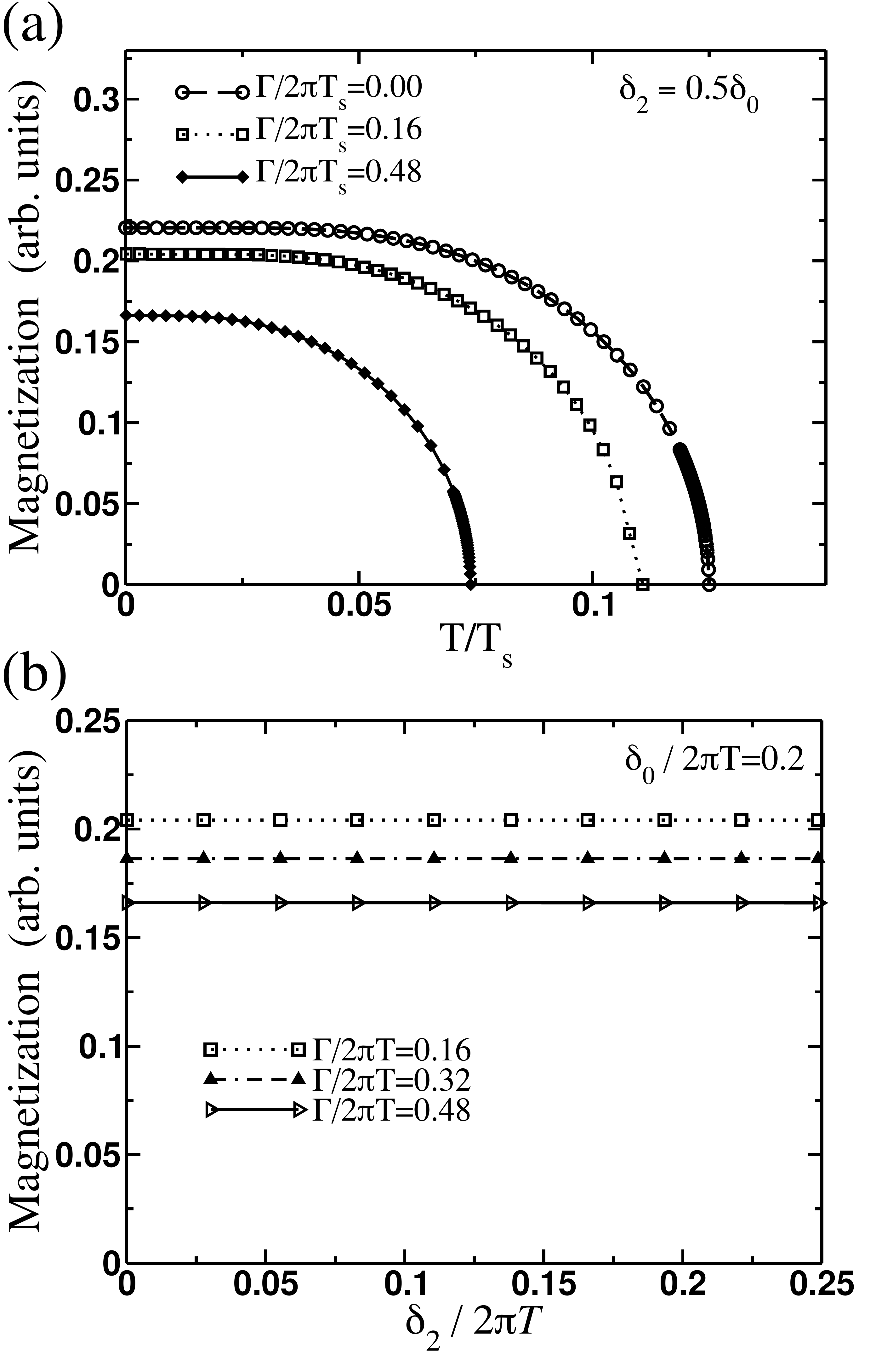}
\caption{Dependence of magnetizations $m_1$ and $m_2$ on temperature and anisotropy parameter $\delta_2$ obtained by the numerical analysis of the self-consistency equations (\ref{selfcons}. Within the numerical accuracy, we found that $m_1=m_2$. Panel (a): magnetization as a function of temperature are plotted for various values of scattering rate $\Gamma$ and $\delta_2=0.5\delta_0$ with $\delta_0=2\pi T_s$ where $T_s$ is a Curie temperature in isotropic system without impurities. Panel (b): magnetization as a function of the Fermi surface anisotropy parameter $\delta_2$ and $T=0.1T_s$.}
\label{Fig1Mag}
\end{figure}

The first two quasiclassical equations (\ref{sol}) can be re-written in a compact form using relations (\ref{constraint}). Indeed, by introducing the auxiliary variables
\beg\label{u1u2}
u_1=\frac{m_X\sqrt{g_1}}{\sqrt{2+g_1+g_2}}, \quad u_2=\frac{m_Y\sqrt{g_2}}{\sqrt{2+g_1+g_2}}, 
\en
the quasiclassical equations acquire the following form
\beg\label{Cubic}
\begin{split}
&u_1\left[2i\Omega_n+\delta_2\cos(2\phi)-u_1-u_2\right]=m_X^2, \\
&u_2\left[2i\Omega_n-\delta_2\cos(2\phi)-u_1-u_2\right]=m_Y^2.
\end{split}
\en
Perhaps, for the clarity of our subsequent discussion  it would be useful to mention that in the case when magnetizations are vanishingly small, $m_{X,Y}\ll \pi T$, functions $g_{1,2}\propto m_{X,Y}^2$, $q_x\propto m_Xm_Y$, while $p_{x,y}\propto m_{X,Y}$.

We have to analyze the solution of the equations (\ref{Cubic}) in two special cases only: (i) single-${\vec Q}$ state for which we set $m_Y=0$ and $m_X=m_1$ and (ii) double-${\vec Q}$ state in which $m_X=m_Y=m_2/\sqrt{2}$. 
\paragraph{Single-${\vec Q}$ state.}  Since in this case $p_y=q_x=g_2=0$, we have
\beg\label{pxQ1}
p_x(i\omega_n,\phi)=\left(\frac{m_1^2+u_1^2}{m_1^2-u_1^2}\right)\frac{2m_1}{2i\Omega_n+\delta_2\cos(2\phi)}.
\en
In turn, function $u_1(i\omega_n,\phi)$ is determined by one of the two roots of the quadratic equation (first equation in (\ref{Cubic}) with $u_2=0$) which recovers the correct expression for the non-interacting propagator:    
\beg\label{u1Q1}
\begin{split}
u_1&=Z_n(\phi)-\gamma\sqrt{Z_n^2(\phi)-m_1^2},
\end{split}
\en
where $Z_n(\phi)=i\Omega_n+(\delta_2/2)\cos(2\phi)$ and $\gamma$ is the prefactor which guarantees that in the limit when $m_1\to 0$, $u_1$ also vanishes. 
\paragraph{Double-${\vec Q}$ state.} The solution of the equations (\ref{Cubic}) in this cases reduces to the solution 
of a single cubic equation
\beg\label{Cubicvt}
(x+2i\Omega_n)[x^2-\delta_2^2\cos^2(2\phi)+m_2^2]=2i\Omega_nm_2^2.
\en
Functions $u_1$ and $u_2$ can then be computed from
\beg\label{Cubicu1u2}
u_{1,2}=\frac{1}{2}\left(1+\frac{2i\Omega_n}{x_a}\right)\left[x_a\pm\delta_2\cos(2\phi)\right],
\en
where $x_a$ is one of the roots of equation (\ref{Cubicvt}). 

It is \emph{a priori} not clear which one of the three roots must be chosen. An additional difficulty in choosing the correct root consists in the fact that after finding an analytic expressions for the roots (\ref{Cubicvt}) it turns out that depending on the limiting case ($m_{X,Y}\to 0$ or $\delta_2\to 0$, for example) different roots recover the correct expressions for the quasiclassical functions. The procedure we have adopted consisted in analyzing all three complex roots of (\ref{Cubicvt}) and picking up the one for which all the equations (\ref{sol},\ref{constraint}) are satisfied and in addition $\textrm{Im}[p_{x,y}]<0$.  
The latter condition guarantees the positive contribution to magnetization, Eq. (\ref{selfcons}), and minimum in free energy.

\paragraph{Results.} We have used thes expressions to evaluate the dependence of the order parameters $m_1$ and $m_2$ on the anisotropy parameter $\delta_2$ for a fixed value of $\delta_0$ and fixed temperature. Naturally, we find that both $m_1$ and $m_2$ are the same for the same values of the model parameters. The results of the calculations for the temperature dependence of the magnetizations $m_1$ and $m_2$ are presented on Fig. \ref{Fig1Mag}(a). Perhaps it is not too surprising that we found the values of $m_1$ and $m_2$ equal to each other within the error bars of the numerical calculations. Therefore, self-consistency equations cannot be used to determine which of the two states would be more favorable and we will have to compute the free energy for each state. 

\subsection{Disordered system}
Quasiclassical equations for the disordered system naturally have similar form as equations (\ref{sol}) for the fact that the matrix structure of the quasiclassical function does not change as soon as $\Gamma$ becomes nonzero. The calculation of the commutation relations (\ref{Quasi1}) yields
\beg\label{Dissol}
\begin{split}
&\left[2i\widetilde{\Omega}_n+\Delta_n(\phi)\right]p_x+(m_X-i\Gamma\av{p_x})(g_2-2g_3)\\&=i\Gamma\av{q_x}p_y-(m_Y-i\Gamma\av{p_y})q_x, \\
&\left[2i\widetilde{\Omega}_n-\Delta_n(\phi)\right]p_y+(m_Y-i\Gamma\av{p_y})(g_1-2g_3)\\&=i\Gamma\av{q_x}p_x-(m_X-i\Gamma\av{p_x})q_x. 
\end{split}
\en
In these equations $\widetilde{\Omega}_n=\Omega_n+(\Gamma/4)(4\av{g_3}-\av{g_1}-\av{g_2})$ and $\Delta_n(\phi)=\delta_2\cos(2\phi)+i(\Gamma/2)(\av{g_1}-\av{g_2})$.
Just like in the case $\Gamma=0$ the third equation is redundant and therefore is not listed here. 

Equations (\ref{Dissol}) show that disorder renormalization plays out differently for single-${\vec Q}$ and double-${\vec Q}$ states. Given these disorder renormalizations, in order to solve the self-consistency equation (\ref{selfcons}), the angular averages ($\av{g_3}$ and $\av{p_x}$ in a single-${\vec Q}$ state, for example) had to be computed by iterations.  We found that the values of the corresponding magnetizations still remain essentially identical for nonzero 
$\Gamma$, Fig. \ref{Fig1Mag}(b). We also found, that qualitative behavior of both $m_1(\delta_2)$ and $m_2(\delta_2)$ does not change with an inclusion of disorder. 

Lastly, we would like to mention that the inclusion of the interband disorder with scattering rate $\Gamma_\pi$ would not change the dependence of the magnetization on the anisotropy parameters, but only leads to a faster suppression of the magnetization with an increase in $\Gamma_\pi$. 
\begin{figure}[h]
\includegraphics[width=8cm]{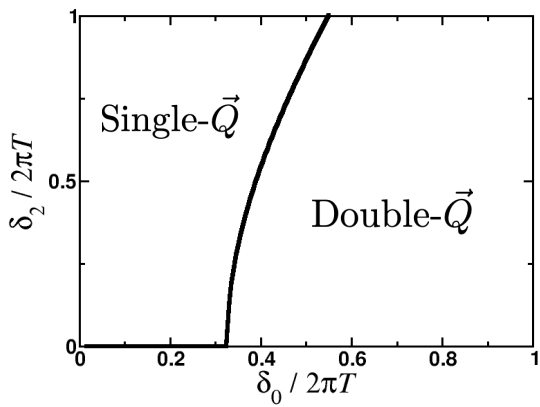}
\caption{Results of the numerical analysis of the coefficient $g_4$ in the free energy expansion for the clean system, 
$\Gamma=0$. The solid line marks the first order transition line along which the coefficient $g_4$ is zero and the energies of the single-${\vec Q}$ and double-${\vec Q}$ states are degenerate. }
\label{FigPhaseDiagram}
\end{figure}

\section{Free energy}
To derive an expression for the free energy in terms of the quasiclassical functions, we can employ an expression for the effective action corresponding to the model Hamiltonian (\ref{Eq1}). Omitting the disorder potential for now, we have \cite{AlexC4}
${\cal F}(m_X,m_Y)=({m_X^2+m_Y^2})/{g_{\textrm{sdw}}}-S(\lambda=1)$ with
\beg\label{Smf}
\begin{split}
&S(\lambda)=T\sum\limits_{i\omega_n}\int_\bk\textrm{Tr}\log\left(\hat{\mathbbm{1}}+\lambda\hat{G}_0(i\omega_n,\bk)\hat{W}\right).
\end{split}
\en
Here $\hat{G}_0(i\omega_n,\bk)$ is the single-particle propagator for the non-interacting system, 
$\hat{W}=- m_X\hat{\cal S}_{X}-m_Y\hat{\cal S}_{Y}$ and
\beg\label{SxSy}
\hat{\cal S}_{X}=\left(
\begin{matrix}
0 & \hat{\sigma}_3 & 0 \\
\hat{\sigma}_3 & 0 & 0  \\
0 & 0 & 0
\end{matrix}
\right), \quad \hat{\cal S}_{Y}=\left(
\begin{matrix}
0 & 0 & \hat{\sigma}_3 \\
0 & 0 & 0  \\
 \hat{\sigma}_3 & 0 & 0
\end{matrix}
\right).
\en

\begin{figure}[h]
\includegraphics[width=7cm]{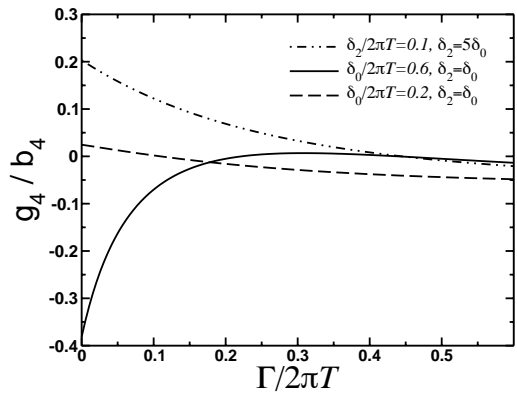}
\caption{Plot of the nematic coupling constant $g_4$ which appears in free energy as a function of the disorder scattering rate for various values of the anisotropy parameters $\delta_0$ and $\delta_4$.}
\label{Fig-g4}
\end{figure}

The expression for the free energy in terms of the quasiclassical functions can be derived by following the steps in the calculation of Ref. \onlinecite{ReinerFree1984}. First, we note
\beg\label{Slam}
\begin{split}
&\frac{\partial S}{\partial\lambda}=-i\pi\nu_F\int\limits_0^{\pi}\frac{d\phi_\bk}{\pi}\textrm{Tr}\left[\hat{\cal M}_3\hat{\cal G}_\lambda(i\omega_n,\phi_\bk)\hat{W}\right],
\end{split}
\en
where $\hat{\cal G}_\lambda(i\omega_n,\phi_\bk)$ is found from solving the quasiclassical equations (\ref{Quasi1}) in which order parameters have been rescaled by parameter $\lambda$,  $m_{X,Y}\to\lambda m_{X,Y}$. The resulting expression for the free energy reads
\beg\label{Free}
\begin{split}
&{\cal F}(m_X,m_Y)=\frac{m_X^2+m_Y^2}{g_{\textrm{sdw}}}\\&-2i\pi\nu_FT\sum\limits_{i\omega_n}\int\limits_0^1d\lambda\left(m_X\av{p_{\lambda x}}+m_Y\av{p_{\lambda y}}\right).
\end{split}
\en
This expression can also be employed for the case of non-zero disorder by using the solution of equations 
(\ref{Dissol}) with the rescaled magnetizations. 

It is a hopeless task to evaluate the free energy (\ref{Free}) analytically, but it is amenable to the numerical analysis. However, our numerical computation of the free energy for the single-$Q$ and double-$Q$ states ran into an unexpected problem: the difference between the free energies of the corresponding states fall within the numerical error of the calculation. Thus, in order to determine which one of the two magnetic states will be energetically favorable, below we derive the Landau expansion. 

\subsection{Free energy expansion in powers of the magnetization}
Having found an expression for the free energy, we consider the temperatures slightly below the critical temperature, so that 
both magnetizations are sufficiently small compared to $\pi T$. Then, we can formally obtain the solution of the 
quasiclassical equations (\ref{Dissol}) by expanding functions $p_{x}$ and $p_y$ in powers of $m_X$ and $m_Y$.
\paragraph{Clean case.} In the case of the clean system the expression up to the fourth order in powers of magnetization reads
\beg\label{F4clean}
\begin{split}
{\cal F}(m_X,m_Y)&=a_2\left(m_X^2+m_Y^2\right)+b_4\left(m_X^2+m_Y^2\right)^2\\&-g_4\left(m_X^2-m_Y^2\right)^2
+O(m^6),
\end{split}
\en
where the corresponding coefficients are given by $b_4=(a_4+a_{XY})/2$, $g_4=(a_{XY}-a_4)/2$ with
\beg\label{CleanCoeffs}
\begin{split}
a_2&=\frac{1}{g_{\textrm{sdw}}}-8\nu_FT\sum\limits_{\omega_n>0}\int\limits_0^{\pi}\frac{\omega_nd\phi_\bk}{4\omega_n^2+(\delta_0\pm\delta_2\cos(2\phi_\bk))^2}, \\ 
a_4&=4\nu_F\textrm{Im}\left\{T\sum\limits_{\omega_n>0}\int\limits_0^{\pi}\frac{d\phi_\bk}{\left[2i\Omega_n\pm\delta_2\cos(2\phi_\bk)\right]^3}\right\}, \\ 
a_{XY}&=-8\nu_F\textrm{Re}\left\{T\sum\limits_{\omega_n>0}\int\limits_0^{\pi}\frac{\Omega_nd\phi_\bk}{\left[4\Omega_n^2+\delta_2^2\cos^2(2\phi_\bk)\right]^2}\right\}.
\end{split}
\nonumber
\en
The sign of the coefficient $c_4$ is crucial for it determines which one of the two states becomes energetically more favorable. Indeed, let us assume that we choose the model parameters such that both $m_1$ and $m_2$ are much smaller than $\pi T$.  For a fixed value of $m_1=m_2$ it follows that when $g_4>0$ the single-${\vec Q}$ will have the lower energy compared to the double-${\vec Q}$ one.
However, one needs to keep in mind that this line of arguments holds only when the coefficients in the free energy expansion are all of the order $O(1)$ and coefficient $b_4$ remains positive for a given set of values of parameters $\delta_0/2\pi T$ and $\delta_2/2\pi T$.
\begin{figure}[h]
\includegraphics[width=7cm]{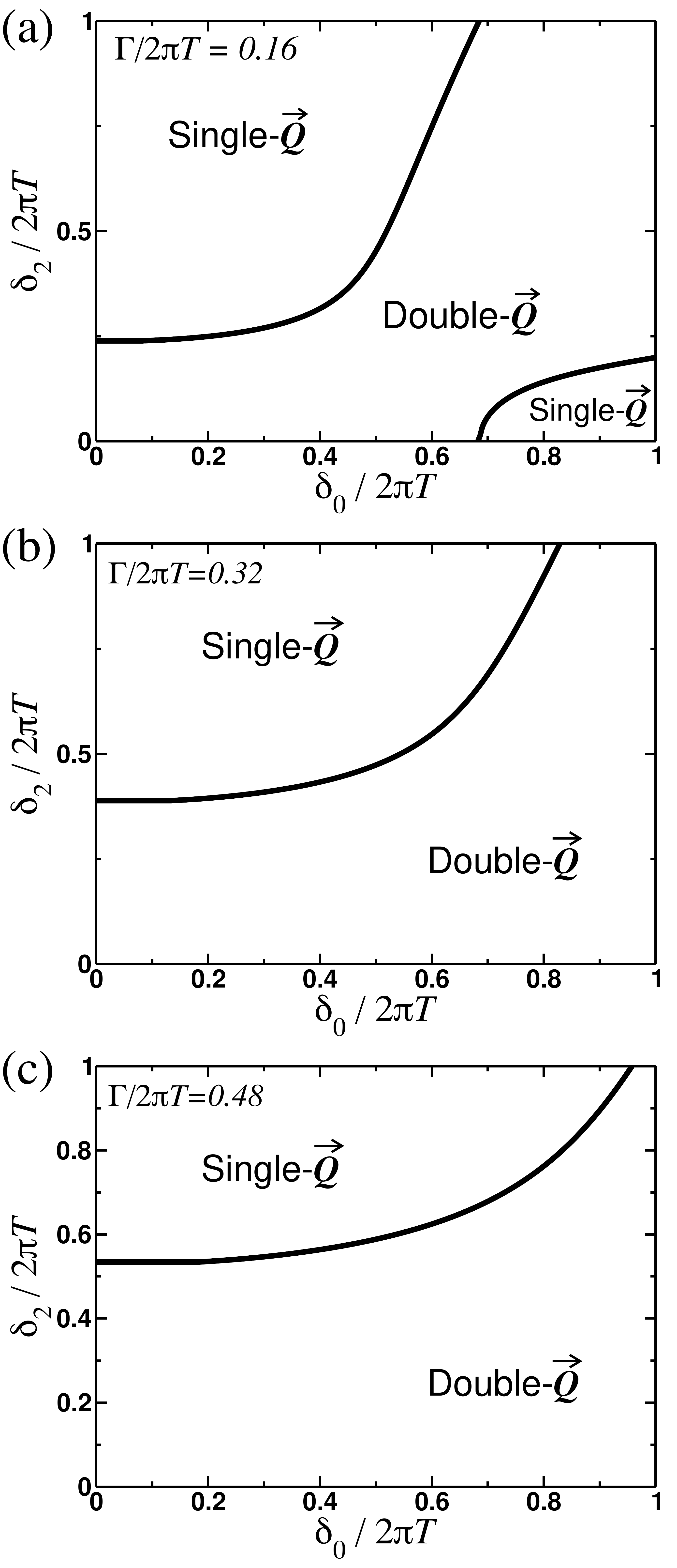}
\caption{Results of the numerical analysis of the coefficient $g_4$ in the free energy expansion for the disordered system.
With the increase in the value of the disorder scattering rate, the single-${\vec Q}$ state is a ground state for higher and higher values of the anisotropy parameter $\delta_2$ which accounts for the ellipticity of the electron-like pockets.}
\label{FigPhaseDiagramDis}
\end{figure}

\paragraph{Disordered case.}
The question arises as to how nonzero disorder will affect the stability of the single-${\vec Q}$ state.\cite{AlexC4}
The calculation of the quasiclassical functions is similar to the one in the clean case, with the only exception that the averages over the angle $\phi$ need to be computed self-consistently. For example, the first order corrections to functions $p_x$ and $p_y$ are
\beg\label{GammaPi0}
\begin{split}
p_x^{(1)}&=\frac{2\left(m_X-i\Gamma\av{p_x^{(1)}}\right)}{2i\left(|\omega_n|+\Gamma\right)\textrm{sign}(\omega_n)+\delta_0+\delta_2\cos(2\phi)}, \\ 
p_y^{(1)}&=\frac{2\left(m_Y-i\Gamma\av{p_y^{(1)}}\right)}{2i\left(|\omega_n|+\Gamma\right)\textrm{sign}(\omega_n)+\delta_0-\delta_2\cos(2\phi)}.
\end{split}
\en
After integrating both parts of these expressions over $\phi$, we can easily solve for $\av{p_x^{(1)}}$ and $\av{p_y^{(1)}}$.
 
The calculation of the expressions for the coefficients of the Landau expansion in this case gives
\beg\label{DisCoeff}
\begin{split}
A_4&=4\nu_F\textrm{Im}\left\{T\sum\limits_{\omega_n>0}\left(\frac{\eta(i\omega_n)-i\Gamma}{\eta(i\omega_n)+i\Gamma}\right)^3\right.\\ &\left.\times\int\limits_0^{\pi}\frac{d\phi_\bk}{\left[2i(\Omega_n+\Gamma)\pm\delta_2\cos(2\phi_\bk)\right]^3}\right\}, \\
A_{\textrm{XY}}&=-8\nu_F\textrm{Re}\left\{T\sum\limits_{\omega_n>0}\left(\frac{\eta(i\omega_n)-i\Gamma}{\eta(i\omega_n)+i\Gamma}\right)^3\right.\\&\left.\times\int\limits_0^{\pi}\frac{(\Omega_n+\Gamma)d\phi_\bk}{\left[4(\Omega_n+\Gamma)^2+\delta_2^2\cos^2(2\phi_\bk)\right]^2}\right\}\\&-2\pi\nu_F\Gamma\textrm{Re}\left\{T\sum\limits_{\omega_n>0}\left(\frac{\eta(i\omega_n)-i\Gamma}{\eta(i\omega_n)+i\Gamma}\right)^3z^2(i\omega_n)\right\}.
\end{split}
\nonumber
\en
Functions $\eta(i\omega_n)$ and $z(i\omega_n)$ appear as a result of disorder renormalization and are listed in Appendix. The coefficient $g_4$ in free energy is now given by 
$g_4=(A_{XY}-A_4)/2$. Compared with the clean case, we see that expression for the coefficient 
$A_{\textrm{XY}}$ contains an extra term proportional to $\Gamma$. The dependence of $g_4$ on disorder can be easily analyzed numerically. The results of the numerical computations are shown in Fig. \ref{Fig-g4}.

\subsection{Phase diagram}
To determine the phase diagram in the space of anisotropy parameters $\delta_0$ and $\delta_2$, we need to 
find a point where the free energies of both states become degenerate, $g_4(\delta_{0c},\delta_{2c})=0$. In Fig. \ref{FigPhaseDiagram} we show the phase diagram for the clean system. It agrees qualitatively with the one obtained previously:\cite{AlexC4} for small values of $\delta_2/2\pi T\ll 1$, single-${\vec Q}$ state becomes energetically favorable when the value of electron-hole asymmetry $\delta_0$ is above a critical value $\delta_{0c}/2\pi T\sim 0.3$. 

With an addition of disorder, phase diagram is modified and the results are presented on Fig. \ref{FigPhaseDiagramDis} For small disorder the critical line separating two phases slightly moves to higher values of $\delta_2$. Perhaps unexpectedly, a small region of single-${\vec Q}$ state appears at large (compared to $\delta_2$) values of $\delta_0$. Upon further increase in the values of the disorder scattering rate, the phase boundary separating two states moves to higher values of $\delta_2$ and also extends to higher values of $\delta_0$. Overall, we may conclude that disorder promotes double-${\vec Q}$ state over the 
single-${\vec Q}$ state. 

\section{Discussion}
As we have already pointed out in the Introduction, our main goal was to demonstrate how the quasi-classical method can be applied to analyse the competition between magnetic states in multiband metals in the presence of disorder. Having accomplished that goal, we can now generalize it to investigate the problem of an interplay between superconductivity and magnetism. It is already well established that by including the interband disorder scattering Anderson-Abrikosov-Gor'kov theorem makes it possible for superconductivity and magnetism to co-exist in a certain region of the phase diagram, which size is determined by the ratio of the intra- and inter-band scattering rates.\cite{Vavilov2011,Dzero2015} The question is then would be to check if superconducting order may provide an additional contribution in determining which of the two competing magnetic states would be energetically favorable. These results may be employed to provide a qualitative understanding as to why nematicity has been observed in stoichiometric iron selenide in contrast to electron-doped iron selenide.

Lastly, we would like to mention that the inclusion of the interband disorder scattering would not affect our results in any substantial way. Indeed, compared to the case of intraband disorder, the inclusion of the interband scattering leads primarily to the faster suppression of the critical temperature, without affecting the ground state energies of the single- and double-${\vec Q}$ states significantly. 

To summarize, in this paper we have formulated the quasi-classical approach to analyze the relative stability of the single- and double-${\vec Q}$ spin-density-wave states with respect to band and effective mass anisotropy as well as disorder scattering. Generally, we find that with an increase in intraband disorder scattering rate, the system favors the single-${\vec Q}$ for moderately high values of the Fermi surface anisotropy parameter, $\delta_2$.

\section{Acknowledgments}
We would like to thank Alex Levchenko for bringing this problem to ou rattention and many fruitful conversations. Useful discussions with R. M. Fernandes and E. K\"{o}nig are gratefully acknowledged. 
This work was financially supported by the U.S. Department of Energy, Basic Energy Sciences, grant DE-SC0016481 (MD) and by the Israel Science Foundation, Grant No. 1287/15 (MK). 

\begin{appendix}
\section{Coefficients in the free energy expansion}
I this Section we provide the details of the calculation for the Landau free energy expansion. Both $p_{\lambda x}$ and $p_{\lambda y}$ can be determined approximately for small values of $m_X$ and $m_Y$ from the quasiclassical equations. We start with the derivation for the clean case, $\Gamma=0$. 
\subsection{First order corrections}
Up to the linear order in $m_j$ from Eqs. (\ref{sol}) I find $g_3^{(0)}=\textrm{sign}(\omega_n)$ and 
\beg\label{FirstOrder}
\begin{split}
p_{\lambda x}^{(1)}&=\frac{2\lambda m_X\textrm{sign}(\omega_n)}{2i\omega_n+\delta_0+\delta_2\cos(2\phi)}, \\ p_{\lambda y}^{(1)}&=\frac{2\lambda m_Y\textrm{sign}(\omega_n)}{2i\omega_n+\delta_0-\delta_2\cos(2\phi)}.
\end{split}
\en
\subsection{Third order corrections}
The second order correction to $p_{\lambda j}$ is zero. To determine the third order correction, we first need to compute the second order corrections to $g_j$'s. To do that, we first use equations (\ref{Cubic}) (and presume for simplicity that $\omega_n>0$):
\beg\label{SecondOrder}
\begin{split}
u_{\lambda 1}^{(2)}&=\frac{\lambda^2m_X^2}{2i\omega_n+\delta_0+\delta_2\cos(2\phi)}, \\
u_{\lambda 2}^{(2)}&=\frac{\lambda^2m_Y^2}{2i\omega_n+\delta_0-\delta_2\cos(2\phi)},
\end{split}
\en
so that
\beg\label{gj2}
\begin{split}
g_{\lambda 1}^{(2)}&=\frac{2(\lambda m_X)^2}{\left[2i\omega_n+\delta_0+\delta_2\cos(2\phi)\right]^2}, \\
g_{\lambda 2}^{(2)}&=\frac{2(\lambda m_Y)^2}{\left[2i\omega_n+\delta_0-\delta_2\cos(2\phi)\right]^2}, \\
g_{\lambda 3}^{(2)}&=g_{\lambda 1}^{(2)}+g_{\lambda 2}^{(2)}.
\end{split}
\en
In addition, for the function $q_x$ we find
\beg\label{qx2}
q_{\lambda x}^{(2)}=-\frac{2\lambda^2m_Xm_Y}{\left(2i\omega_n+\delta_0\right)^2-\delta_2^2\cos^2(2\phi)}.
\en
The choice of sign follows from considering the trivial case of $\delta_2=0$.

Given all these expressions, we go back to equations (\ref{sol}) to obtain the following expression:
\beg\label{px3}
\begin{split}
p_{\lambda x}^{(3)}&=\frac{4(\lambda m_X)^3}{\left[2i\Omega_n+\delta_2\cos(2\phi)\right]^3}\\&+\frac{8i\lambda^3\Omega_nm_Xm_Y^2}{\left[2i\Omega_n+\delta_2\cos(2\phi)\right]^2\left[2i\Omega_n-\delta_2\cos(2\phi)\right]^2}.
\end{split}
\en
Similarly, for $p_{\lambda y}^{(3)}$ I find
\beg\label{px3}
\begin{split}
p_{\lambda y}^{(3)}&=\frac{4(\lambda m_Y)^3}{\left[2i\Omega_n-\delta_2\cos(2\phi)\right]^3}\\&+\frac{8i\lambda^3\Omega_nm_Ym_X^2}{\left[2i\Omega_n+\delta_2\cos(2\phi)\right]^2\left[2i\Omega_n-\delta_2\cos(2\phi)\right]^2}.
\end{split}
\en
After plugging these expressions into Eq. (\ref{Free}) and grouping the similar terms, we arrive to Eq. (\ref{F4clean}).

\begin{widetext}
\subsection{Functions $\eta(i\omega_n)$ and $z(i\omega_n)$}
The formulas for the coefficients in Landau free energy expansion (\ref{DisCoeff}) include the following functions:
\beg\label{eta}
{\eta^{-1}(i\omega_n)}=\frac{1}{\pi}\int\limits_0^\pi\frac{d\phi}{2i\left(\Omega_n+\Gamma\right)\pm\delta_2\cos(2\phi)}, \quad
z(i\omega_n,\delta_0)=-\frac{1}{\pi}\int\limits_0^\pi\frac{d\phi}{4(\Omega_n+\Gamma)^2+\delta_2^2\cos^2(2\phi)}.
\en

\end{widetext}
\end{appendix}

\bibliography{elinembib}

\end{document}